\documentclass[12pt]{article}
\usepackage[pdftex]{graphicx}
\usepackage{epsfig}
\usepackage{epsf}
\newcommand{\xiit}{\tilde{\mbox{\boldmath $\xi$}}^{(i)}}

\newcommand{\KLd}{Kullback-Leibler divergence~}

\newcommand{\Ef}{{\rm E}_f}
\newcommand{\EKL}{{\rm EKL}}
\newcommand{\KL}{{\rm KL}}
\newcommand{\AIC}{{\rm AIC}}
\newcommand{\Tr}{{\rm Tr}}
\newcommand{\tendD}{\longrightarrow \hspace{-0.50cm} ^D \hspace{0.50cm}}
\newcommand{\N}{{{\cal N}}}
\newcommand{\dpg}[2]{\frac{\partial #1}{\partial #2}}
\newcommand{\ddpg}[2]{\frac{\partial^2 #1}{\partial #2^2}}
\newtheorem{Definition}{Definition}

\newtheorem{Lemma}{Lemma}

\begin{document}
\title{Estimating a difference of Kullback-Leibler risks using a normalized difference of AIC}

 \author{D. Commenges $^{1,2}$, A. Sayyareh $^{1,3}$, L. Letenneur $^{1,2}$, \\J. Guedj $^{1,2}$ and A. Bar-Hen $^{5,6}$}
 \maketitle

1 INSERM, Epidemiology and Biostatistics Research Center, Bordeaux, F-33076, France

2 University of Bordeaux 2, Bordeaux, F33076, France

3 University Razi, Kermanshah, Iran

5 University Paris Descartes, Paris, F-75270, France

6 CNRS, UMR 8145, Paris, F-75270, France

 E-mail: daniel.commenges@isped.u-bordeaux2.fr


 \newpage
 \begin{center}
{\bf Estimating a difference between Kullback-Leibler risks by a normalized difference of AIC}
 \end{center}

\vspace{\baselineskip}
\begin{center} SUMMARY \end{center}
AIC is commonly used for model selection but the precise value of AIC has no direct interpretation. We are interested in quantifying a difference of risks between two models. This may be useful for both an explanatory point of view or for prediction, where a simpler model may be preferred if it does nearly as well as a more complex model. The difference of risks can be interpreted by linking the risks with relative errors in the computation of probabilities and looking at the values obtained for simple models. A scale of values going from negligible to large is proposed.
We propose a normalization of a difference of Akaike criteria for estimating the difference of expected Kullback-Leibler risks between maximum likelihood estimators of the distribution in two different models. The variability of this statistic can be estimated. Thus, an interval can be constructed which contains the true difference of expected Kullback-Leibler risks with a pre-specified probability. A simulation study shows that the method works and it is illustrated on two examples. The first is a study of the relationship between body-mass index and depression in elderly people.  The second is the choice between models of HIV dynamics, where one model makes the distinction between activated CD4+ T lymphocytes and the other does not.

\vspace{10mm}

\noindent
 {\em Some key words} : Akaike criterion, body-mass index, depression, HIV dynamics, Kullback-Leibler, logistic regression, model choice.


\section{Introduction}

Since its proposal by Akaike (1973), Akaike information criterion (AIC) has had a huge impact on so-called ``model choice'', in particular in the application of statistical methods; see the presentation of deLeuwe (1992). It is often used in its original simple form, precisely because of its simplicity. Many variants of the criterion have been proposed. We may cite in particular the EIC (Konishi and Kitagawa, 1996; Shibata, 1997) which makes use of the bootstrap, extended to the choice of semi-parametric estimators by Liquet, Sakarovitch and Commenges (2004). Other criteria have been proposed such as the BIC (Schwartz, 1978) or approaches based on complexity (Bozdogan, 2000).
AIC is commonly used to select the ``best'' model on the basis of a sample and it is often forgotten that it is a statistic and as such has a distribution (see Burnham and Anderson, 2002, and Shimodaira, 2001). When the goal is prediction or estimating a parameter which may be common to several models, the model averaging approach (Hoeting et al., 1999; Hjort and Claesken, 2003; Shen and Huang, 2006) may be used.

One problem with AIC is that its value has no intrinsic meaning; in particular AIC is not invariant to a one-to-one transformation of the random variables and values of AIC depend on the number of observations. Investigators commonly display big numbers, only the last digits of which are used to decide which is the smallest.  If the specific structure of the models is of interest, because it tells us something about the explanation of the observed phenomena, it may be interesting to measure how far from the truth each model is. This may not be possible but we can quantify the difference of risks between two models. It may also be useful in prediction problems where we may prefer a simpler model, not only on statistical grounds but because of its very simplicity, if the increase of risk incurred by using it is not too large. Of course estimating the difference of risks will be informative only if we have an idea of what a large or a small difference is.

  We show that a normalized difference of AIC is an estimate of a difference of Kullback-Leibler risks. The distribution of this statistic can be estimated using the results of Vuong (1989) for non-nested models and results of Wald (1943) for the case of nested models. We give some examples of values of such differences to help develop an intuition of what a large or a small difference is. 

In section 2 we present two examples. One is the comparison of a linear and a non-linear effect of body-mass index (BMI) on depression  using data from the Paquid study; the other is the comparison of two models of interaction between HIV and the immune system.
In section 3 we present the relevant Kullback-Leibler risk and we show that the normalized difference of AIC is an estimate of the difference of risks; moreover we propose a so-called ``tracking interval'' which should contain the difference of risks with a given probability; we also give insight in the interpretation of the differences of risks. Section 4 presents a simulation study in the framework of the logistic regression, which makes it possible to assess the properties of the proposed tracking interval. In section 5 we present an illustration on real data in the two examples.

\section{Motivating examples}
\subsection{Comparison of linear and non-linear effect models of BMI on depression}
Our first example bears on the comparison of  possible models of association
of depression and Body-mass index (BMI) in elderly people, using the data of the Paquid study (Letenneur et al., 1999). We aim at assessing quantitatively the difference between estimators based on different models.

  As is conventional, depression was considered as a binary trait coded by a dichotomized
version of the CESD (using the thresholds 17 and 23 for men and women respectively). The question here is to see whether there is a linear effect or if there is an optimal BMI, as far as depression is concerned. This problem is treated in the logistic regression framework. The simplicity of the problem makes it possible to design a simulation study  which looks like this real data problem. 

 We worked with the sample
of the first visit of the Paquid study and we excluded the subjects who were diagnosed demented at that visit:
the sample size was 3484. We fitted logistic regression models for explaining depression from BMI, age and
gender. We entered age, gender and their interaction as explanatory variables. As for BMI which was the factor
of main interest, we tried a linear (in the logistic scale) model and then we challenged the linear model by
trying a categorization of BMI in terciles and a quadratic model. Specifically it is interesting to see, if there is an effect of BMI, whether there is a linear trend or there is an optimal region of values of the BMI (as far as depression is concerned). We also tried a more complex model involving simple powers of weight and height.

\subsection{Comparison of two models of interaction between HIV and the immune system}\label{HIV-descr}
Models of the interaction between HIV and the immune system have had a high impact on the research in the pathology induced by HIV (Ho et al., 1995, Perelson et al., 1996). These models are based on ODE systems reflecting the mechanisms of infection of CD4+ T Lymphocytes (called CD4 for short) and the production of viruses by infected cells. A possible model, denoted ${\cal M}_1$, is graphically represented in Figure \ref{figHIV} (a); see Appendix for the description of the system of ordinary differential equations (ODE). Rather than making a patient-by-patient analysis, random effect models (Putter et al., 2002) make it possible to analyze a sample of subjects, thus yielding more precise estimates of the parameters. The statistical estimation in these models is challenging because (i) the ODE systems have no analytical solution; (ii) computation of the likelihood involves numerical multiple integrals.

 It may be useful to distinguish between quiescent and activated CD4 because it seems that only activated CD4 can be infected (De Boer and Perelson, 1998). Guedj, Commenges and Thi\'ebaut (2007) analyzed such a model, denoted ${\cal M}_2$, represented in Figure \ref{figHIV}  (b); see Appendix for details. However this model is more complex and therefore numerically more challenging. Moreover only the total number of CD4 is measured. So one may wonder whether the possible gain obtained with this model is worth the additional complexity. One way to study it is to estimate the difference of Kullback-Leibler risks between the two models. 
Bortz and Nelson (2006) used an information complexity criterion and AIC to select between HIV dynamics models but could not quantitatively assess the difference between models. We will attempt to estimate the difference of Kullback-Leibler risks between ${\cal M}_1$ and ${\cal M}_2$ using data of a clinical trial. 

\section{Theory about inference of differences of AIC criteria}
\subsection{Estimating a difference of Kullback-Leibler divergences}
Consider a sample of independently identically distributed (iid) random variables $\bar Y_n=(Y_i,i=1\ldots,n)$
having probability density function (pdf) $f=f(.)$. Let us consider two models : $(g)=(g^{\beta}(.))_{\beta \in
B}, B\subset \Re^p$ and $(h)=(h^{\gamma}(.))_{\gamma \in \Gamma},\Gamma \subset \Re^q$.
\begin{Definition} (i) $(g)$ and $(h)$ are non-overlapping if $(g) \cap (h)=\emptyset$; (ii) $(g)$ is nested in $(h)$ if  $(g) \subset (h)$;
(iii) $(g)$ is well specified if there is a value $\beta_*\in B$ such that $g^{\beta_*}=f$; otherwise it is misspecified.
\end{Definition}

The log-likelihood loss of $g^\beta$ relatively to $f$ for observation $Y$ is $\log \frac{f(Y)}{g^{\beta}(Y)}$.
The expectation of this loss under $f$, or
risk, is the Kullback-Leibler divergence (Kullback, 1968) between $g^\beta$ and $f$: $\KL(g^{\beta},f)={\rm
E}_f[\log \frac{f(Y)}{g^{\beta}(Y)}]$. We have $\KL(g^{\beta},f)\ge 0$ and $\KL(g^{\beta},f)= 0$ implies that
$g^{\beta}=f$, that is $\beta=\beta_*$.  The Kullback-Leibler divergence is often intuitively interpreted as a
distance between the two pdf (or more generally between the two probability measures) but this is not
mathematically a distance; in particular the \KLd is not symmetric. It may be felt that this is a drawback, and
in particular it makes any graphical representation perilous. However this feature may also have a deep meaning
in our particular problem: there is no symmetry between $f$, the true pdf, and $g^{\beta}$, a possible pdf. So
we shall take on the fact that the \KLd is an expected loss (with respect to $f$) and not a distance. We assume
that there is a value $\beta_0 \in B$ which minimizes $\KL(g^{\beta},f)$. If the model is well specified
$\beta_0=\beta_*$; if the model is misspecified $\KL(g^{\beta_0},f)>0$. The MLE $\hat \beta_n$ is a consistent estimator of $\beta_0$.


We shall say that $(g)$ is {\em closer} to $f$ than $(h)$ (avoiding to qualify $(g)$ as``better'' which may be misleading in this context) if $\KL(g^{\beta_0},f)<\KL(h^{\gamma_0},f)$. We have $\KL(g^{\beta},f)={\rm E}_f[\log {f(Y)}]-{\rm E}_f[\log g^{\beta}(Y)]$. We cannot estimate $\KL(g^{\beta_0},f)$ because the entropy of $f$, $H(f)={\rm E}_f[\log {f(Y)}]$, cannot be correctly estimated. However, we can estimate the difference of risks $\Delta(g^{\beta_0},h^{ \gamma_0})=\KL(g^{\beta_0},f)-\KL(h^{\gamma_0},f)$, a quantitative measure of the difference of misspecification by $-n^{-1}(L^{g^{\hat \beta_n}}_{\bar Y_n}-L^{h^{\hat \gamma_n}}_{\bar Y_n})$.

This result may not be completely satisfactory in practice if $n$ is not very large because the distribution we will use is $g^{\hat \beta_n}$ rather than  $g^{\beta_0}$. Thus it is more relevant to consider the risk $\Ef[\log \frac{f(Y)}{g^{\hat \beta_n}(Y)}]$ that we call the expected Kullback-Leibler risk (or simply Kullback-Leibler risk) and that we denote by $\EKL(g^{\hat \beta_n},f)$. This is the point of view introduced by Akaike (1973).

Akaike's approach was revisited by Linhart and Zucchini (1986) who showed that:
\begin{equation} \label{PP1} \EKL(g^{\hat \beta_n},f)=\KL(g^{\beta_0},f)+\frac{1}{2} n^{-1}\Tr (I_g^{-1}J_g) +o(n^{-1}), \end{equation}
where $I_g=-\Ef[\ddpg{\log g^{\beta}(Y)}{\beta}|_{\beta_0}]$ and $J_g=\Ef\{[\dpg{\log g^{\beta}(Y)}{\beta}|_{\beta_0}][\dpg{\log g^{\beta}(Y)}{\beta}|_{\beta_0}]^T\}$.
This can be nicely interpreted by saying that the risk $\EKL(g^{\hat \beta_n},f)$ is the sum of the misspecification risk $\KL(g^{\beta_0},f)$ plus the statistical risk $\frac{1}{2} n^{-1}\Tr (I_g^{-1}J_g)$. Note in passing that if $(g)$ is well specified we have $\KL(g^{\beta_0},f)=0$ and  $I_g=J_g$, and thus $\EKL(g^{\hat \beta_n},f)=\frac{p}{2n}+o(n^{-1})$.

We also have:
\begin{equation} \label{PP2} \EKL(g^{\hat \beta_n},f)=-\Ef(n^{-1}L^{g^{\hat \beta_n}}_{\bar Y_n})+H(f)+\frac{1}{n} \Tr (I_g^{-1}J_g) +o_p(n^{-1}). \end{equation}
Here we have essentially estimated ${\rm E}_f[\log g^{\beta_0}(Y)]$ by ${\rm E}_f[n^{-1}L^{g^{\hat \beta_n}}]$ but because of the overestimation bias, the factor $\frac{1}{2}$ in the last term disappears; thus the term $\frac{1}{n} \Tr (I_g^{-1}J_g)$ is the sum of two equal terms, the statistical error and the estimation bias of the misspecification risk (of course the misspecification risk is estimated up to the constant $H(f)$).
Akaike criterion ($\AIC(g^{\hat \beta_n})=-2L^{g^{\hat \beta_n}}_{\bar Y_n}+2p$) follows from (\ref{PP2}) by multiplying by $2n$, deleting the constant term $H(f)$  replacing $\Ef(n^{-1}L^{g^{\hat \beta_n}}_{\bar Y_n})$ by $n^{-1}L^{g^{\hat \beta_n}}_{\bar Y_n}$ and replacing $\Tr (I_g^{-1}J_g)$ by $p$.

What we really want to estimate is $\Delta(g^{\hat \beta_n},h^{\hat \gamma_n})=\EKL(g^{\hat \beta_n},f)-\EKL(h^{\hat \gamma_n},f)$.
Using (\ref{PP2}) we obtain:
 $$\Ef \left \{-n^{-1}\{L^{g^{\hat \beta_n}}_{\bar Y_n}-L^{h^{\hat \gamma_n}}_{\bar Y_n}-[\Tr (I_g^{-1}J_g)-\Tr (I_h^{-1}J_h)]\}\right \}=\Delta(g^{\hat \beta_n},h^{\hat \gamma_n})+o_p(n^{-1}).$$
Using the Akaike approximation $\Tr (I_g^{-1}J_g) \approx p$, we obtain a simple estimator of $\Delta(g^{\hat \beta_n},h^{\hat \gamma_n})$:
\begin{equation} \label{D} D(g^{\hat \beta_n},h^{\hat \gamma_n})=\frac{1}{2}n^{-1}[\AIC(g^{\hat \beta_n})-\AIC(h^{\hat \gamma_n})]=-n^{-1}[L^{g^{\hat \beta_n}}_{\bar Y_n}-L^{h^{\hat \gamma_n}}_{\bar Y_n}-(p-q)].\end{equation}
$\Ef [D(g^{\hat \beta_n},h^{\hat \gamma_n})-\Delta(g^{\hat \beta_n},h^{\hat \gamma_n})]$ is an $o(n^{-1})$.
Thus, in contrast with AIC, $D(g^{\hat \beta_n},h^{\hat \gamma_n})$ has an interpretation since its expectation tracks the
quantity of main interest $\Delta(g^{\hat \beta_n},h^{\hat \gamma_n})$ with pretty good accuracy. Moreover it
has important invariance properties.

\begin{Lemma} [Invariance properties] Both $\Delta(g^{\hat \beta_n},h^{\hat \gamma_n})$ and $D(g^{\hat \beta_n},h^{\hat \gamma_n})$ are invariant under re-parametrization, one-to-one transformation of the observed variables and change of the reference probability.
\end{Lemma}

The proof is straightforward. It can be noted that AIC itself is invariant under re-parametrization but neither under one-to-one transformation of the observed variables nor change of the reference probability.

\subsection{Tracking interval for a difference of Kullback-Leibler divergences}
We propose a ``tracking interval'' for $\Delta(g^{\hat \beta_n},h^{\hat \gamma_n})$. This is not a usual confidence interval  because $\Delta(g^{\hat \beta_n},h^{\hat \gamma_n})$ changes with $n$. Although it converges toward $\Delta(g^{\beta_0},h^{\gamma_0})$ we wish to approach $\Delta(g^{\hat \beta_n},h^{\hat \gamma_n})$ for values of $n$ for which the Akaike correction is not negligible.

We focus on the case where $g^{\beta_0} \ne h^{\gamma_0}$. Using Theorem 3.3 of Vuong (1989), which is valid under conditions clearly stated by this author, we obtain that in that case:
\begin{equation} \label{asympt} n^{1/2}[D(g^{\hat \beta_n},h^{\hat \gamma_n})-\Delta(g^{\hat \beta_n},h^{\hat \gamma_n})] \tendD \N (0,\omega_*^2),\end{equation}
where $\omega_*^2={\rm var} \left [\log \frac{g^{\beta_0}(Y)}{h^{\gamma_0}(Y)}\right]$.
A natural estimator of $\omega_*^2$ is
$$\hat \omega_n^2=n^{-1}\sum_{i=1}^n \left [\log \frac{g^{\hat \beta_n}(Y_i)}{h^{\hat \gamma_n}(Y_i)}\right]^2-\left [n^{-1}\sum_{i=1}^n \log \frac{g^{\hat \beta_n}(Y_i)}{h^{\hat \gamma_n}(Y_i)}\right ]^2.$$

From this we can compute the tracking interval $(A_n,B_n)$, where $A_n=D(g^{\hat \beta_n},h^{\hat \gamma_n})-z_{\alpha/2}n^{-1/2}\hat \omega_n$ and $B_n=D(g^{\hat \beta_n},h^{\hat \gamma_n})+z_{\alpha/2}n^{-1/2}\hat \omega_n$, where $1-\Phi(z_{\alpha/2})=\alpha/2$ and $\Phi$ is the cdf of the standard normal variable. This interval has the property:
$$P_f[A_n<\Delta(g^{\hat \beta_n},h^{\hat \gamma_n})<B_n] \longrightarrow 1-\alpha,$$
where $P_f$ represents the probability with density $f$.  The assumption $g^{\beta_0} \ne h^{\gamma_0}$ is necessarily the case if the models do not overlap and may also be often the case even if the models overlap or are nested. However in the latter case the convergence toward the normal may be slow and it is desirable to construct confidence and tracking intervals compatible with the likelihood ratio test.

\subsection{The case of nested models}\label{LR}
In the case of nested models $(g) \subset (h)$ the likelihood ratio test is often used to test whether the true distribution $f$ is in $(g)$. It can be used in the more general case where $(h)$ (and hence $(g)$) is misspecified. In that case the null hypothesis $H_0$ that can be tested by the Likelihood ratio test is $g^{\beta_0}=h^{\gamma_0}$; that is, the closest distribution to $f$ in $(h)$ is in $(g)$. Let us define $LR= L^{g^{\hat \beta_n}}_{\bar Y_n}-L^{h^{\hat \gamma_n}}_{\bar Y_n}$. The asymptotic distribution of $2LR$ under the null hypothesis is Chi-square with $q-p$ degrees of freedom. If $H_0$ is true we have $\KL(g^{\beta_0},f)=\KL(h^{\gamma_0},f)$ and we deduce from (\ref{PP1}) that $\Delta(g^{\hat \beta_n},h^{\hat \gamma_n})\approx \frac{p-q}{2n} <0$. Thus if $H_0$ is true the risk of $g^{\hat \beta_n}$ is always lower than that of $h^{\hat \gamma_n}$, so we should work with $(g)$.

If however $H_0$ is not true we have $KL(h^{\gamma_0}) < KL(g^{\beta_0})$ so that 
\begin{equation} \label{bound} \Delta(g^{\hat \beta_n},h^{\hat \gamma_n}) > \frac{p-q}{2n}.\end{equation} 
Since $\frac{p-q}{2n}$ is negative it is possible, if the difference of misspecification risks is small enough, that the risk incurred with $(g)$ is smaller than that incurred with $(h)$.
Also, if $H_0$ is not true, the LR statistic has a completely different asymptotic distribution than when $H_0$ is true. This is a normal rather than a Chi-square distribution, and even more important, there is a scaling factor $n^{-1/2}$ (see (\ref{asympt})), showing that the LR statistic is an $O_p(n^{1/2})$ and no longer an $O_p(1)$. A practical question arises: is there a transition between two so different distributions ?
When $H_0$ is not true but we are not far from it, that is $|\Delta(h^{\gamma_0},g^{\beta_0})|$ is small, the convergence toward the normal may be slow, so at finite distance we may be in between the chi-square and the normal. In particular we know that $D > (p-q)/n$; a normal distribution giving non-negligible probability to $\{D<(p-q)/n\}$ would not be satisfactory. 

Wald (1943), see also Kendall and Stuart (1973), showed that under the alternative hypothesis, the likelihood ratio statistic ($-2LR$) has approximately a non-central chi-squared distribution with $q-p$ degrees of freedom (dof). We adopt this distribution and express the non-centrality parameter $\delta$ in term of $\Delta(g^{\beta_0},h^{\gamma_0})$. We deduce from equations (1) and (3) that ${\rm E}[-2LR]\approx 2n\Delta(g^{\beta_0},h^{\gamma_0})+q-p$. Since the expectation of a non-central chisquare with $dof=q-p$ is $\delta+ q-p$ we obtain $\delta\approx 2n\Delta(g^{\beta_0},h^{\gamma_0})$. For $\Delta(g^{\beta_0},h^{\gamma_0})=0$ we retrieve the $\chi^2_{q-p}$ distribution for the classical  test of the null hypothesis using the likelihood ratio statistic. This distribution is also compatible with the asymptotic normal distribution given by Vuong (1989). Indeed, for fixed $\Delta(g^{\beta_0},h^{\gamma_0})$, we have $\delta \rightarrow \infty$ when $n\rightarrow \infty$, and we know that the non-central chi-squared distribution tends to a normal when $\delta \rightarrow \infty$ (Evans, Hastings and Peacock, 1993). This also entails that, for fixed $n$, the normal approximation will be better for large $\Delta(g^{\beta_0},h^{\gamma_0})$.

Now suppose that we wish to test ``$\Delta(g^{\beta_0},h^{\gamma_0})=\Delta_0$''. We are in the ideal situation of simple hypothesis testing where we can apply the Neyman Lemma. That is, the rejection region of the test is formed by all the values having the lower values of the density of the test statistic. Typically the rejection region will be $(c,\infty)$  (resp. $(c_{inf},c_{sup})$) for small (resp. large) values of $\Delta_0$. The test can be inverted to form a confidence interval for $\Delta(g^{\beta_0},h^{\gamma_0})$: the $1-\alpha$ confidence interval is formed of all the values $\Delta_0$ which are not rejected by the test at level $\alpha$. This confidence interval is by definition compatible with the likelihood ratio test, since $0$ will not be in the interval if  ``$\Delta(g^{\beta_0},h^{\gamma_0})=0$'' has been rejected by the test (which precisely assumes a $\chi^2_{q-p}$ distribution for $\Delta_0=0$). From this confidence interval for $\Delta(g^{\beta_0},h^{\gamma_0})$, say $(A'_n,B'_n)$,  we can deduce the tracking interval for $\Delta(g^{\hat \beta_n},h^{\hat \gamma_n})$ by subtracting to the bounds the additional statistical risk incurred with $(h)$, that is $(q-p)/2n$: $A_n=A'_n+(p-q)/2n$; $B_n=B'_n+(p-q)/2n$. It is not impossible that $A_n$ be negative, even if ``$\Delta(g^{\beta_0},h^{\gamma_0})=0$'' has been rejected. Indeed, if we reject $H_0$ using the likelihood ratio test, we reject $\Delta(g^{\hat \beta_n},h^{\hat \gamma_n}) =\frac{p-q}{2n}$ but we do not reject negative values of $\Delta(g^{\hat \beta_n},h^{\hat \gamma_n})$ larger than $\frac{p-q}{2n}$. 

In practice, the computation of the intervals may be done by computing the p-value for each value $\Delta_0$. Let $f_{\Delta_0}$ and $F_{\Delta_0}$ be the pdf and cdf of the non-central chi-squared distribution with $q-p$ dof and non-centrality parameter $2n\Delta_0$. If $f_{\Delta_0}(x)>f_{\Delta_0}(-2LR)$ for all $ x<-2LR$,  the p-value is simply $1-F_{\Delta_0}(-2LR)$. This situation occurs for small values of dof and non-centrality parameter. If this is not  the case the rejection region includes an interval $(0,c_{inf})$ so the p-value is $1-F_{\Delta_0}(-2LR) + F_{\Delta_0}(c_{inf})$ where $f_{\Delta_0}(c_{inf})=f_{\Delta_0}(-2LR)$. In practice it may not be easy to find $c_{inf}$ unless a special program is available. We propose to look at the quantile of $(1-F_{\Delta_0}(-2LR))/2$, say $q_{pv/2}$. If $f_{\Delta_0}(q_{pv/2})>f_{\Delta_0}(-2LR)$ we can take p-value$=1-F_{\Delta_0}(-2LR)$; if $f_{\Delta_0}(q_{pv/2})<f_{\Delta_0}(-2LR)$ we  take p-value$=2(1-F_{\Delta_0}(-2LR))$.

\subsection{How to interpret a difference of Kullback-Leibler risks}

It is important to judge whether the values within the intervals correspond to large or small expected losses. The Kullback-Leibler risk takes values between $0$ and $+\infty$ but in practice most of the risks or difference of risks that we encounter are lower than $1$. To give an idea of how to interpret these values we may relate them to relative errors made in evaluation of probabilities as in Commenges et al. (2007). We will make errors by evaluating the probability of an event $A$ using a distribution $g$, $P_g(A)$, rather than using the true distribution $f$, $P_f(A)$. For instance we may evaluate the relative error $r_e( P_g(A), P_f(A))=\frac{P_f(A)- P_g(A)}{P_f(A)}$. Consider the typical event on which $P_f(A)$ will be under-evaluated defined as:
$A=\{x: g(x)<f(x)\}$. To obtain a simple formula relating $KL(g, f)$ to the error on $P_f(A)$ we consider the particular case $P_f(A)=1/2$ and $g/f$ constant on $A$ and $A^C$. In that case we easily find: $r_e(P_g(A), P_f(A))=\sqrt{1-e^{-2KL(g, f)}}\approx \sqrt{2KL(g, f)}$, the approximation being valid for small $KL$ value.
For $KL$ values of  $10^{-4}$, $10^{-3}$, $10^{-2}$, $10^{-1}$ we find that $r_e( P_g(A), P_f(A))$ is equal to $0.014$, $0.045$, $0.14$ and $0.44$, errors that we may qualify as ``negligible'',  ``small'',``moderate'' and ``large'' respectively.

 As already noted we can give an
interpretation of $\EKL$ from (\ref{PP1}) as the sum of the misspecification risk $\KL(g^{\beta_0},f)$ and the estimation risk, approximated by $p/2n$. For a well specified model the risk is about $p/2n$; for instance it is $10^{-2}$ if $p=10$ and $n=500$, or if $p=1$ and $n=50$. The statistical risk associated to the estimation of one parameter is negligible, small, moderate and large for $n= 5000, 500, 50, 5$ respectively. The correspondence between the different scales is summarized in Table 1. 
We may also measure on this scale the magnitude of the Akaike correction of $(p-q)/n$.

 As an example the $\KL$ divergence of a double exponential
relative to a normal  distribution with same mean and variance is of order $10^{-1}$ what may be called a
``large'' value. As another example we may compute the risk incurred when using a normal distribution of variance $\sigma^2$ when the true distribution has variance one. It is easy to compute that the Kullback-Leibler risk is $\frac{1}{2} [\log \sigma ^2 -1 +\frac{1}{\sigma^2}]$: this expression takes the value $0$ for $\sigma^2=1$ and tends toward $+\infty$  if $\sigma^2$ tends toward $+\infty$ or $0$. The values obtained for $\sigma^2= 1.02; 1.1; 1.3; 2$ are respectively $= 0.0001; 0.002; 0.016; 0.096$ corresponding approximately to the the negligible, small, moderate and large levels. To approach a risk of $1$, one has to take very large values of $\sigma^2$: the risk is $0.65$ for $\sigma^2=4$ and $0.91$ for $\sigma^2=16$. Finally we give the correspondence between the KL divergence and the odds-ratio in a particular case of a binary variable with $P_f(Y=1|X)=1/2$, while ${\rm logit} [P_g(Y=1|X)]=\beta X$, $X$ being itself a binary variable taking values $1$ or $-1$ with probability $1/2$. We have $KL(g, f) = {\rm E} \{1/2\log [\frac{1/2}{P_g(Y=1|X)}+1/2\log [\frac{1/2}{P_g(Y=0|X)}\}$, where the expectation bears on $X$. After some algebra we find that $KL(g, f) = 1/2 \log [1/2(1+\cosh (\beta))]$. The values of the odds-ratio (OR$=e^{\beta}$) giving negligible, small, moderate and large divergences are $1.03; 1.1; 1.35; 2.5$ respectively. It is important to realize that this correspondence depends on the joint distribution of both $Y$ and $X$; higher values of OR are associated to the same divergence levels for $P_f(Y=1|X) \neq 1/2$ or $P(X=1) \neq 1/2$.

A question which arises is whether the Kullback-Leibler risks are comparable when $Y$ is multivariate and when $Y$ is univariate. If we have $n$ independent univariate variables and we group them in vectors of size $m$, we obtain $n'=n/m$ multivariate observations. To get the same estimator of the difference of risks between two models we should divide by the $n'm$ rather than by $n'$. Thus in case of multivariate data we propose to divide the difference of AIC by the total number of measurements to get a value that is more comparable to situation where the variables are univariate. 

\subsection{Extension to regression models}
All that has been said can be extended to regression models $(g_{Y|X})=(g_{Y|X}^{\beta}(.|.))_{\beta \in B}$ and $(h_{Y|X})=(h_{Y|X}^{\gamma}(.|.))_{\gamma \in \Gamma}$. This can be done as in Vuong (1989) by directly defining the \KLd in term of conditional densities: $\KL(g_{Y|X}^{\beta},f_{Y|X})={\Ef}[\log \frac{f_{Y|X}(Y|X)}{ g_{Y|X}^{\beta}(Y|X)}]$, where the expectation is taken for the true distribution of the couple $Y,X$. However this approach has the drawback of requiring a new definition of the \KLd. The so-called reduced model approach (Commenges et al., 2007) is more satisfactory.
Consider a sample of iid couples of variables $(Y_i,X_i), i=1,\ldots,n$ having joint pdf $f$, $f(y,x)=f_{Y|X}(y|x)f_X(x)$. Consider the model $(g)=(g^{\beta}(.,.))_{\beta \in B}$ such that $g^{\beta}(y,x)=g^{\beta}_{Y|X}(y|x)f_X(x)$ ; the model is called ``reduced'' because $f_X(.)$ is assumed known. The \KLd is:
$$\KL(g^{\beta},f)=\Ef [\log f_{Y|X}(Y|X)]-\Ef [\log g^{\beta}_{Y|X}(Y,X)],$$
that is the term in $f_X(.)$ disappears (so that we do not need to know it in fact) and we get the same
definition as in Vuong (1989) using only the conventional \KLd.

\section{Simulation study}
\subsection{Study of the tracking interval in a non-nested case}
We performed a simulation resembling the situation of the Depression-BMI application where we have to choose between different logistic regression models. We considered iid samples of size $n$ of triples $(Y_i,x_1^i,x^i_2), i=1,\ldots,n$ from the following distribution (which plays the role of the true distribution $f$). The conditional distribution of $Y_i$ given $(x_1^i,x^i_2)$ was logistic with ${\rm logit} [f_{Y|X}(1|x_1^i,x^i_2)]=0.5+x_1^i+2x_2^i$, where $f_{Y|X}(1|x_1^i,x^i_2)=P_f(Y_i=1|x_1^i,x^i_2)$; the marginal distributions of $(x_1^i,x^i_2)$ were bivariate normal with zero expectation and variance equal to the identity matrix. We considered model $(g)$ specified by ${\rm logit} [g^{\beta}_{Y|X}(1|x_1^i,x^i_2)]=\beta_0+\beta_1x_1^i+\beta_2x_2^i$, which was well specified and the (mis)specified model $(h)$ defined as ${\rm logit} [h^{\gamma}_{Y|X}(1|x_1^i,x^i_2)]=\gamma_0+ \sum_{l=1}^2 \gamma_l x^i_{1l}+\gamma_{3}x_2^i$, where $x^i_{1l}$ were dummy variables indicating in which categories $x^i_1$ fell; the categories were defined using terciles of the observed distribution of $x_1$, and this was represented by two dummy variables: $x^i_{11}$ indicating whether $x^i_1$ fell in the first tercile or not, $x^i_{12}$ indicating whether $x^i_1$ fell in the second tercile or not.

Since model (g) is well specified we know that $g^{\beta_0}=f$, that the misspecification error $\KL(g^{\beta_0},f)$ is zero and that $ \Tr (I_g^{-1} J_g)=p$. As for model (h) we must compute the quantities of interest by simulation. We can compute that in the logistic regression the $l,k$ term of the matrix $J_h$ is $\Ef[x_l(Y-\frac{e^{x\gamma_0}}{1+e^{x\gamma_0}})^2x_k]$, and that the  $l,k$ term of the matrix $I_h$ is $\Ef[x_l\frac{e^{x\gamma_0}}{(1+e^{x\gamma_0})^2}x_k]$. We estimated $\gamma_0$ by fitting model $(h)$ on a simulated data set with $n=10^5$. Our precise estimate $\check \gamma_0$ was thus $\hat \gamma_n$ for $n={10^5}$. We used it to precisely estimate $J_h$ and $I_h$ as $\check I_h=10^{-5}\sum_{i=1}^{10^5}[x^i_l\frac{e^{x^i\check \gamma_0}}{(1+e^{x^i\check \gamma_0})^2}x^i_k]$ and $\check J_h=10^{-5}\sum_{i=1}^{10^5}[x^i_l(Y_i-\frac{e^{x^i\check \gamma_0}}{1+e^{x^i\check \gamma_0}})^2x^i_k]$.

 We estimated $\KL(h^{\gamma_0},f)$ by $10^{-5}\sum_{i=1}^{10^5} \log \frac{f_{Y|X}(Y_i|x_1^i,x^i_2)}{h^{\check \gamma_0}_{Y|X}(Y_i|x_1^i,x^i_2)}$.  We also computed a precise estimate of $\omega_*^2$, $\check \omega_*^2$, by the empirical variance of $\log \frac{f_{Y|X}(Y_i|x_1^i,x^i_2)}{h^{\check \gamma_0}_{Y|X}(Y_i|x_1^i,x^i_2)}$ computed on $10^5$ replicas. Thus we can compute a precise estimate of $\EKL(h^{\hat \gamma_n},f)$ and $\EKL(g^{\hat \beta_n},f)$ by replacing the terms on right-hand of (\ref{PP1}) by their estimates. Because $(g)$ is well specified we obtain immediately $\EKL(g^{\hat \beta_n},f)\approx \frac{3}{2n}$; a precise estimate of $\EKL(g^{\hat \beta_n},f)-\EKL(h^{\hat \gamma_n},f)$ is thus given by $\check \Delta=\frac{3}{2n}-\KL(h^{\check \gamma_0},f)-\frac{1}{2n} \Tr (\check I_h^{-1}\check J_h)$. We find first that $\KL(h^{\check \gamma_0},f)\approx 7.28~ 10^{-3}$, a value approaching the ``moderate magnitude''. We found 3.998 and 3.999 for the values of $\Tr (\check I_h^{-1}\check J_h)$ for $n=250$ and $n=1000$ respectively. These values are very close to $q=4$ (that would obtain if $(h)$ was well-specified) so, in the following we will use this approximation. Using this approximation we can compute $\check \Delta=-\frac{1}{2n}-\KL(h^{\check \gamma_0},f)$ and obtain  $\check \Delta= -9.28~ 10^{-3}$ for $n=250$ and $\check \Delta= -7.78~ 10^{-3}$ for $n=1000$. We also find $\check \omega_*^2=1.44~ 10^{-2}$. We can then compute the standard error of $D$ as $n^{-1/2}\check \omega_*$ and find $7.59~ 10^{-3}$ and $3.79~ 10^{-3}$ for $n=250$ and $n=1000$ respectively. We see at once that there is more chance that the tracking interval does not contain zero for $n=1000$ than for $n=250$.

We generated 1000 replications from the above model for $n=250$ and $n=1000$. For each replication we computed the maximum likelihood estimates and the AIC. We computed the histogram of $D(g^{\hat \beta_n},h^{\hat \gamma_n})$ (see Figure 2): its shape is approximately in accordance with the asymptotic normal distribution for both sample sizes; the empirical mean was $-9.50~ 10^{-3}$ and $-7.67 ~10^{-3}$ for $n=250$ and $n=1000$ respectively, close to the values of $\check \Delta$. The empirical variance of $D$ (not shown) was in agreement with the theoretical variance computed from $\check \omega_*^2$. The mean of the estimated variances $\hat \omega_*^2$ was $1.88  ~10^{-2} $ and $1.54  ~10^{-2}$ for $n=250$ and $n=1000$ respectively, also reasonably close to the $\check \omega_*^2$. The proportion of replicas for which $\check \Delta$ was outside the $.95$ tracking interval was $0.045$ and $0.053 $ for $n=250$ and $n=1000$ respectively. The proportion of replicas for which zero was outside of the tracking interval was $0.197$ and $0.514$ for $n=250$ and $n=1000$ respectively, and in all cases $(g)$ was preferred to $(h)$. These results are summarized in Table 2.

The results of the simulation are in accordance with the asymptotic theory. From a practical point of view, the variability of $D$ seems to be large so that it is difficult to be sure that an estimator is better than another one if the difference of risk is small or moderate. Note that this variability is not specific to our approach but is a fact applying to any criteria based on likelihood ratio. For instance in the simulated situation for $n=250$ there is a probability of about 12\% that $D(g^{\hat \beta_n},h^{\hat \gamma_n})$ takes a positive value (thus suggesting the wrong choice) and this probability is exactly the same for AIC.

\subsection{Quality of the fit by the non-central chi-squared distribution in the nested case}
We performed another simulation for the case of nested model, to check the quality of the approximation of the distribution of $-2LR$ by the non-central chi-squared distribution. We made two simulations with true distributions $f^1$, specified by : ${\rm logit}[f^1_{Y|X}(1|x_1^i,x^i_2)]=0.5+0.2 x_1^i+2x_2^i$ and $f^2$, specified by:
 ${\rm logit}[f^2_{Y|X}(1|x_1^i,x^i_2)])=0.5+0.5 x_1^i+2x_2^i$. For both cases we considered two models: $(g)$ and $(h)$ with 
${\rm logit} [g^{\beta}_{Y|X}(1|x_1^i,x^i_2)]=\beta_0+\beta_2x_2^i$ and ${\rm logit} [h^{\gamma}_{Y|X}(1|x_1^i,x^i_2)]=\gamma_0+\gamma_1x_1^i+\gamma_2x_2^i$, so that $(h)$ was well specified while $(g) \subset (h)$ was misspecified. However if $f^1$ is the true distribution the difference of risks using $(g)$ and $(h)$ is of ``small'' magnitude ($\approx 10^{-3}$) while if $f^2$ is the true distribution it of ``moderate'' ($\approx 10^{-2}$) magnitude. The distributions of $(x_1^i,x^i_2)$ were as in the first simulation above. We simulated 10000 replications of samples of size $n=1000$ from  $f^1$ and $f^2$ and in both cases we studied the fit of the non-central chi-squared distribution for the distribution of $-2LR$. The dof was equal to $1$ and we took the expectation equal to the mean, from which we deduced the non-centrality parameter. Figure 3 displays the histograms and the non-central chi-squared densities for both cases. The fits are nearly perfect and we also see that the distribution is closer to the normal for $f^2$ than for $f^1$. It is clear that the convergence to the normal is slow in the case of nested models unless the difference of risks is large.

\section{Applications}
\subsection{Relation between BMI and depression: analysis of the Paquid data}

The values of AIC, and the $D$ statistic and tracking intervals (taking as reference the linear model) are given
in Table 3. The tercile model had a larger AIC than the linear model but the point estimate (D) of the
difference of risks was lower than $10^{-4}$ a level that we have qualified ``negligible'', and zero was well
inside the tracking interval. So from the point of view of Kullback-Leibler risk there was no evidence that one
model is better than the other. When it comes to comparing the linear and the quadratic model, because the first
is nested in the second, we can use the likelihood ratio test: the null hypothesis is that the best distribution
is in the linear sub-model. The hypothesis was strongly rejected ($p < 0.01$). We tend to conclude that the
shape of the effect is not linear and that we may approach it better with a quadratic term. However it is interesting to estimate the difference of risks between the two models. The point estimate
of the difference of risks was $0.0007$, a value which approaches the $10^{-3}$ level that we qualified to be a
small (but not negligible) difference. Since $(g) \subset (h)$ we computed the tracking interval applying the version of the tracking interval for nested models of section 3.3. The computation was done using using the pchisq, dchisq and qchisq R functions. We found $(0.00012; 0.0030)$ for the confidence interval of $\Delta(g^{\beta_0},h^{\gamma_0})$ and, subtracting the increased statistical risk $(p-q)/2n=0.00014$, we found $(-0.00002; 0.0029)$ for the tracking interval. Thus we are not completely sure to incur a smaller risk  with the quadratic model. However, if the difference of risks was not in favor of the quadratic model, this would be completely negligible.  The difference of risks in favor of the quadratic model may be negligible or of small magnitude.

In conclusion there is no reason to prefer
the tercile model to the linear model but there are some reasons to prefer the quadratic model to the linear
model. Figure 4 shows the shape of the effect of BMI with the quadratic model, taking as reference the median
BMI (equal to 24.2). This is a U-shaped curve yielding the lower risks of depression for medium values of the
BMI, somewhat shifted however toward large BMI. Of course the epidemiological interpretation of this result is
delicate and the apparent effect that we have detected is the consequence of complex biological and
psychological mechanisms that we do not attempt to explore here. Several other studies have found links between
BMI and depression (Bergdahl et al., 2007; Bjerkeset et al., 2008).

Since BMI is a combination of weight and height one may wonder whether it is possible to find a better model directly using simple powers of height and weight in the linear predictor. It happens that the model including weight, height, weight$^2$, height$^2$ and $1/$height, that we denote $(w)=(w^{\theta})_{\theta\in \Theta}$, has a better AIC than the quadratic (in BMI) model, $(h)$. Note that $(h)$ is not nested in $(w)$. Following the conventional use of AIC we should prefer $(w)$ to $(h)$. However $(w)$ lacks readability because it involves a combination of weight and height that has never been used. For instance a nice graphical representation of the effect of weight and height such as presented in Figure 4 is not possible. So we have non-statistical reasons to prefer $(h)$ over $(w)$. If we examine the statistical reasons to prefer $(w)$ over $(h)$ they are very thin. First, the point estimate of $\Delta(h^{\hat \beta_n},w^{\hat \theta_n})$ is $D=0.0003$, of the ``negligible'' order of magnitude. Second, the tracking interval is $[-0.0016; 0.0022]$: zero is well inside this interval, so there is no confidence that we incur a lower risk using $(w)$ rather than $(h)$. Thus it is reasonable to prefer $(h)$ for further use, for instance presentation of the epidemiological evidence of a relation between over- and under-weight and depression.

\subsection{Interaction between HIV and the immune system: analysis of the ALBI data}
As an application of the proposed method, we analyzed
the difference of risks between the model ${\cal M}_1$ and model ${\cal M}_2$ described in section \ref{HIV-descr} using the data of  a randomized clinical trial, the ALBI ANRS 070
trial (Molina et al., 1999). This trial compared over 24 weeks the combination of zidovudine plus
lamivudine (AZT+3TC) to that of stavudine plus didanosine
(ddI+d4T). There were 50 patients in each arm.  Measurements of CD4 and of HIV RNA were taken once a month up to six months. The likelihood, taking into account the detection limit of HIV RNA, was computed with the algorithm of Guedj, Thi\'ebaut and Commenges (2007). The AIC for model ${\cal M}_1$ was equal to 1466.15 while for model ${\cal M}_2$ AIC =  1026.63. The estimate of the variance was $\hat \omega^2_n=5.88$. Thus the $D$ statistic was equal to $4.40$. However this applies to a multivariate outcome: we had seven measurements of viral load and of CD4 counts for each subject, that is 14 measurements per subjects. So the standardized value of $D$ was $4.40/14= 0.31$. For the tracking interval we find $[0.28; 0.35]$.

 We can say with a good degree of confidence that the difference of risks is larger than $0.28$, a large difference as we have seen. This means that this difference between quiescent and activated CD4 is an important biological fact and that it must be taken into account, even though fitting the more complicated model is more challenging. 

\section{Discussion}

We have proposed a statistic which tracks the difference of expected Kullback-Leibler risks between maximum likelihood estimators in two different models, $\Delta(g^{\hat \beta_n},h^{\hat \gamma_n})$. Moreover we have an estimator of the variance of this statistic and we can construct a ``tracking interval''. We can also construct a confidence interval for $\Delta(g^{\beta_0},h^{\gamma_0})$: the bounds of the latter are the bounds of the former shifted of $(q-p)/2n$.  The results of our simulation study were in agreement with the asymptotic results. Our approach enlightens the unavoidable variability of any criterion based on log-likelihood ratio such as AIC, BIC and their variants. This variability is generally not taken into account and there is a misleading intuition that extrapolates the distribution of the likelihood ratio test to the variability of AIC. The distribution of the likelihood ratio statistic is well approximated by a normal in the non-nested case while it is better approximated by a non-central chi-squared in the nested case. In both cases the variance is larger than that of the chi-squared with $q-p$ dof, a distribution which holds only under the null hypothesis of the likelihood ratio test.

{\em In fine} we can do more than simply choosing the estimator which has the lowest AIC. We can estimate the difference of risks and this has the same meaning in different problems. We may become accustomed to considering  differences of $10^{-4}$, $10^{-3}$, $10^{-2}$, $10^{-1}$  as negligible, small, moderate and large respectively, as we are accustomed to interpret correlation coefficients or odds-ratios for instance. More work is needed however to deepen our intuition about the magnitude of a difference of Kullback-Leibler risks.


In the first application we have found that the quadratic model for the effect of BMI on risk of depression was better than a linear model, although the difference between the two models was small. With the quadratic model both low and high BMI are at higher risk of depression. Our method gives arguments to prefer the quadratic model in BMI for presentation of the results to a more complex model obtaining a slightly better AIC. In the application on comparing two HIV dynamics models, we found that the model distinguishing quiescent and activated CD4 was better than the simpler model which did not make this distinction. The estimated difference of risks was large and this has implications in future developments of HIV dynamics models. 

The statistic $D$ and the tracking interval for the difference of risks are easy to compute and could be useful in a wide variety of applications.

\setlength{\parindent}{0.0in}
\begin{center}
{\bf REFERENCES}
\end{center}

Akaike, H. (1973). Information theory and an extension of maximum likelihood principle, Second International Symposium on Information Theory, Akademia Kiado, 267-281.\vspace{3mm}


Bergdahl, E, Allard, P., Lundman, B. and Gustafson, Y. ( 2007). Depression in the oldest old in urban and rural municipalities. {\em Aging \& Mental Health}, {\bf 5}: 570-578.\vspace{3mm}

Bjerkeset, O., Romundstad, P.,  Evans, J. and Gunnell, D. (2008). Association of Adult Body Mass Index and Height with Anxiety, Depression,  and Suicide in the General Population: The HUNT Study. {\em Am. J. Epidemiol.} {\bf 167}: 193-202. \vspace{3mm}

Bortz, D. M.  and Nelson,P. W. (2006). Model Selection and Mixed-Effects Modeling of HIV Infection Dynamics. {\bf 68}, 2005-2025.\vspace{3mm}

Bozdogan, H. (2000). Akaike's information criterion and recent developments in information complexity. {\em J. Math. Psych.} {\bf 44}, 62-91.\vspace{3mm}

Burnham, K.P. and Anderson D.R. (2002). {\em Model Selection and Multi-Model Inference: a practical information-theoretic approach}, New-York: Springer.\vspace{3mm}

Commenges, D., Joly, P, G\'egout-Petit, A. and Liquet, B. (2007). Choice between semi-parametric estimators of Markov and non-Markov multi-state models from generally coarsened observations. {\em Scandinavian Journal of Statistics}, {\bf 34}, 33-52.\vspace{3mm}

De Boer, R. and  Perelson, A. S. (1998). Target cell limited and immune control models of HIV infection: a
comparison. {\em J. Theor. Biol.} {\bf 190}, 201-214.\vspace{3mm}

deLeuwe, J. (1992). Introduction to Akaike (1973) Information theory and an extension of the maximum likelihood principle, in {\em Breakthroughs in Statistics} (Kotz, S. and Johnson, N.L., Eds), New-York: Springer.\vspace{3mm}

Evans, M., Hastings, N. and Peacock, B. (1993). {\em Statistical distributions}, Wiley.\vspace{3mm}

Guedj, J., Thi\'ebaut, R. and Commenges, D. (2007) Maximum Likelihood Estimation in Dynamical Models of HIV. {\em Biometrics}, {\bf 63}, 1198-1206.\vspace{3mm}

Hjort, N.L. and Claeskens, G. (2003). Frequentist model average estimators. {\em Journal of the American Statistical Association} {\bf 98}, 879-899.\vspace{3mm}

Ho, D.D., Neumann, A.U., Perelson, A.S., Chen, W., Leonard, J.M. and Markowitz, M.(1995) Rapid turnover of plasma virions and CD4 lymphocytes in HIV-1 infection. {\em Nature} 1995, 373(6510), 123-126.\vspace{3mm}

Hoeting, J.A., Madigan, D., Raftery, A.E. and Volinsky, C.T. (1999). Bayesian model averaging: a tutorial. {\em Statistical Science} {\bf 14}, 332-417.\vspace{3mm}

Kendall, M.G. and Stuart A. (1973). {\em The advanced Theory of Statistics}. London, Charles Griffins. \vspace{3mm}

Konishi, S. and Kitagawa, G. (1996). Generalised information criteria in model selection. {\em Biometrika} {\bf 83}, 875-890\vspace{3mm}

Kullback, S. (1968). {\em Information Theory and Statistics}, New York: Dover.\vspace{3mm}

Letenneur, L., Gilleron, V., Commenges, D., Helmer, C., Orgogozo, JM. and Dartigues, JF. (1999). Are sex and educational level independent predictors of dementia and Alzheimer's disease ? Incidence data from the PAQUID project. {\em  Journal of Neurology Neurosurgery and Psychiatry} {\bf 66}, 177-183.\vspace{3mm}

Linhart, H. and Zucchini,W. (1986). {\em Model Selection}, New York: Wiley.\vspace{3mm}

Liquet, B., Sakarovitch, C. and Commenges, D. (2003).
Bootstrap choice of estimators in parametric and semi-parametric families: an extension of EIC {\em Biometrics} {\bf 59}, 172-178.\vspace{3mm}

Molina, J., Ch\^ene, G., Ferchal, F., Journot, V., Pellegrin, I.,
Sombardier, M. N., Rancinan, C., Cotte, L., Madelaine,
I., Debord, T., and Decazes, J. M. (1999). The ALBI
Trial: A randomized controlled trial comparing stavudine
plus didanosine with zidovudine plus lamivudine and
a regimen alternating both combinations in previously
untreated patients infected with human immunodeficiency
virus. {\em The Journal of Infectious Diseases} {\bf 180}, 351-358. \vspace{3mm}

Perelson, A.S., Neuman, A.U., Markowitch, M., Leonard, J.M. and Ho, D.D. (1996). HIV-1 dynamics in vivo: virion clearance rate, infected cell life-span, and viral generation time. {\em Science}, {\bf 271}, 1582-1586.\vspace{3mm}

Putter, H., Heisterkamp, S. H., Lange, J. M. A. and deWolf, F. (2002). A Bayesian approach to parameter estimation in HIV dynamic models. {\em Stat. Med.}, {\bf 21}, 2199-2214. \vspace{3mm}


Shen, X. and Huang, H-C. (2006). Optimal model assessment, selection and combination. {\em J. Am. Statist. Assoc.} {\bf 101}, 554-568.\vspace{3mm}

Schwarz, G. (1978). Estimating the dimension of a model, {\em Ann. Statist.} {\bf 6}, 461-464.\vspace{3mm}

Shibata, R. (1997). Bootstrap estimate of Kullback-Leibler information for model selection, {\em Statist. Sin.} {\bf 7}, 375-394. \vspace{3mm}

Shimodaira, H. (2001). Multiple Comparisons Of Log-Likelihoods And Combining Nonnested Models With Applications To Phylogenetic Tree Selection. {\em Commun. Statist. Theory Methods} {\bf 30}, 1751, 1772.\vspace{3mm}

Vuong, Q.H. (1989). Likelihood Ratio Tests for Model Selection and Non-Nested Hypotheses.
{\em Econometrica} {\bf 57}, 307-333\vspace{3mm}

Wald, A. (1943). Tests of Statistical Hypotheses Concerning Several Parameters When the Number of Observations is Large. {\em Transactions of the American Mathematical Society} {\bf 54}, 426-482.

\section*{Appendix: The HIV dynamics models}

To write the differential equation for the model, one uses assumptions which are plausible in view of the knowledge of the biological mechanisms: for instance we assume that new CD4 are produced (by the thymus) at a rate $\lambda$, that only activated cells can be infected, that the probability of meeting of a cell and a virion is proportional to the product of their concentrations. A possible model (${\cal M}_1$) takes into account the uninfected and infected CD4, $\bar T$ and $T^*$ respectively, and the viral particles, $V$ and is as follows: 

\begin{eqnarray*}\label{(1)}
d\bar T_t&=& (\lambda - (1-\eta I^{RT}  )\gamma T_tV_{t} - \mu_{\bar T}\bar T_t)dt \\
dT^{*}_t&= &[(1-\eta I^{RT}  )\gamma  T_tV_{t} - \mu _{T^{*}} T^{*}_t]dt \\
dV_{t}& =& (\mu _{T^{*}_t}\pi T^{*}_t - \mu _{v}V_{t})dt, \\
\end{eqnarray*}

where $I^{RT}$ is  indicates whether a treatment based on an inhibitor of the reverse transcriptase.

Another model (${\cal M}_2$) distinguishes between quiescent ($Q$) and activated ($T$) CD4:

\begin{eqnarray*}\label{(2)}
dQ_t&=& (\lambda + \rho T_t - \alpha Q_t- \mu_{Q} Q_t)dt \\
dT_t&=& (\alpha Q_t  - (1-\eta I^{RT}  )\gamma T_tV_{t} - \rho T_t - \mu_{T}T_t)dt \\
dT^{*}_t&= &[(1-\eta I^{RT} )\gamma  T_tV_{t} - \mu _{T^{*}} T^{*}_t]dt \\
dV_{t}& =& ( \mu _{T^{*}_t}\pi T^{*}_t - \mu _{v}V_{t})dt \\
\end{eqnarray*}

A statistical model is necessary to take into account that some parameters may differ from one subject to another and to link the observations to the ODE system. In model ${\cal M}_1$ the parameters $\lambda$ and $\pi$ were random (adding other random parameters did not increase the likelihood). In model ${\cal M}_2$ the parameters $\alpha$, $\lambda$ and $\mu_{T^*}$ were considered as random.  Measurements of the total numbers of CD4 and of number of viruses were available at times $t_{ij}$. We assumed the following observation equations: 

\begin{eqnarray*}\label{observation2}
Y_{ij1}&=& log_{10}(V_I(t_{ij},\xiit) +V_{NI} (t_{ij},\xiit) ) + \epsilon_{ij1} ,~~j \le n_{i} \\
Y_{ij2}&=& (Q(t_{ij},\xiit) +T (t_{ij},\xiit)+T^{*} (t_{ij},\xiit) )^{0.25} + \epsilon_{ij2},~~j \le n_{i} 
\end{eqnarray*}

An additional complexity was that HIV RNA load was measured up to a detection limit. 
Guedj, Thi\'ebaut and Commenges (2007) designed a special algorithm for computing and maximizing likelihood for this type of models. We refer the reader to this paper for more details.
\newpage
\begin{table}
\caption{Order of magnitude of KL risks; the relative error is that for a typical underestimated event in a standard case; the sample size is the size which gives the corresponding statistical risk for estimating one parameter.}
\vspace{10mm}
\begin {center}

\begin{tabular}{c|c|c|c}
\hline
Qualification & KL scale & Relative error & Risk for estimation of one parameter\\
 & & &  Sample size \\
\hline
Large & $10^{-1}$ & $0.44$ & $5$ \\
Moderate & $10^{-2}$ & $0.14$ & $50$ \\
Small & $10^{-3}$ & $0.045$ & $500$ \\
Negligible & $10^{-4}$ & $0.014$ & $5000$ \\
\hline
\end{tabular}
\end{center}

\end{table}

\newpage

\begin{table}
\caption{Simulation study: choice between tercile and linear model for the explanatory variable in a logistic regression model.}
\vspace{10mm}
\begin {center}

\begin{tabular}{c|ccccc}
\hline
n & $\check \Delta$ & $\bar D$ & $\bar {\hat \omega}^2$ & Coverage rate & Power\\
\hline
250 & $-9.28~10^{-3}$ &$-9.50~10^{-3}$& $1.88~10^{-2}$ & $0.967$ & $0.197$\\
1000 & $-7.78~10^{-3}$ &$-7.67~10^{-3}$& $1.54~10^{-2}$ & $0.954$ & $0.514$\\

\hline
\end{tabular}
\end{center}

\end{table}

\newpage
\begin{table}
\caption{Upper part of the table: comparison of the linear, tercile and quadratic models for the effect of BMI on depression: $D$ and the tracking interval are with respect to the linear model. Lower part: comparison of the quadratic model with the model $(w)$ including  weight, height, weight$^2$, height$^2$ and $1/$height: $D$ and the tracking interval are with respect to the quadratic model.}
\vspace{10mm}
\begin {center}

\begin{tabular}{c|ccccc}
\hline
Model & \# parameters & Likelihood &  AIC & $D$ & Tracking interval\\
\hline
\\

Linear & $5$ & $-1346.2$ & $2702.5$ & - & - \\
Tercile & $6$ & $-1345.6$ & $2703.2$ & $-0.0001$ & $[-0.0009; 0.0007]$\\
quadratic & $6$ & $-1342.9$  & $2697.9$ & $0.0007$ & $[-2.10^{-5}; 0.0029]$\\
\hline
quadratic & $6$ & $-1342.9$  & $2697.9$ & - & - \\
$(w)$ & 9 & $-1338.7$ & $2695.5$ & $0.0003$ & $[-0.0016; 0.0022]$ \\
\hline
\end{tabular}
\end{center}

\end{table}
\newpage

\begin{figure}[ht] 
\begin{center}
\includegraphics[scale=0.5,angle=90]{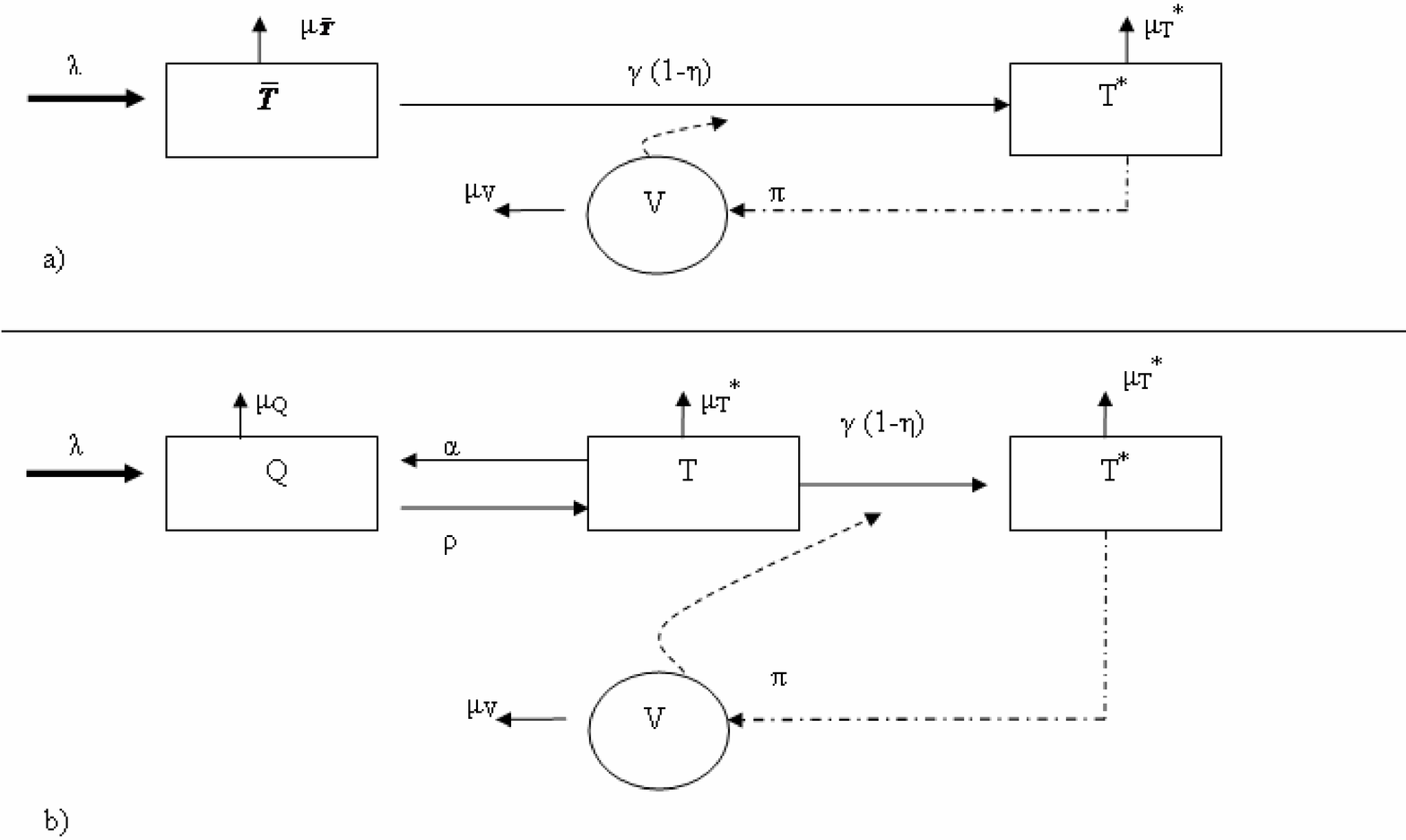}
  \caption{Graphical representation of HIV dynamics models: (a) model ${\cal M}_1$ including uninfected ($\bar T$) and infected ($T^*$) CD4+ T lymphocytes, and HIV viruses ($V$); (b) model ${\cal M}_2$ including uninfected quiescent ($Q$), uninfected activated ($T$), infected ($T^*$) CD4+ T lymphocytes, and HIV viruses ($V$). }
  \label{figHIV}
\end{center}
\end{figure}

\newpage

\begin{figure}[ht]
\begin{center}
\includegraphics[scale=0.8]{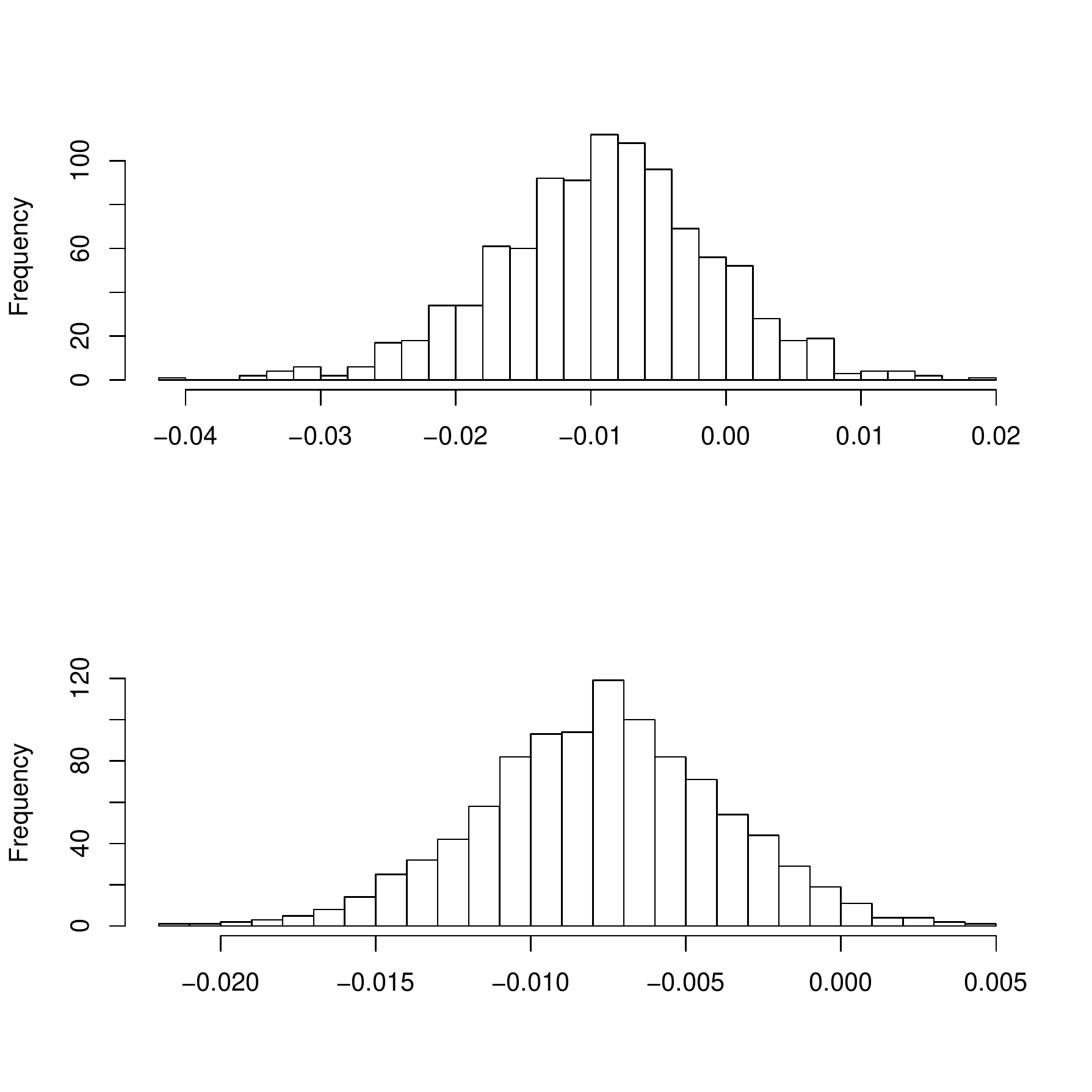}
  \caption{Histogram of the values of $D$ (which estimates the difference of Kullback-Leibler risks between the tercile and the linear models) in the simulation: upper figure, $n=250$, lower figure, $n=1000$. }
  \label{Histog}
\end{center}
\end{figure}
\newpage

\begin{figure}[ht]
\begin{center}
\includegraphics[scale=0.8]{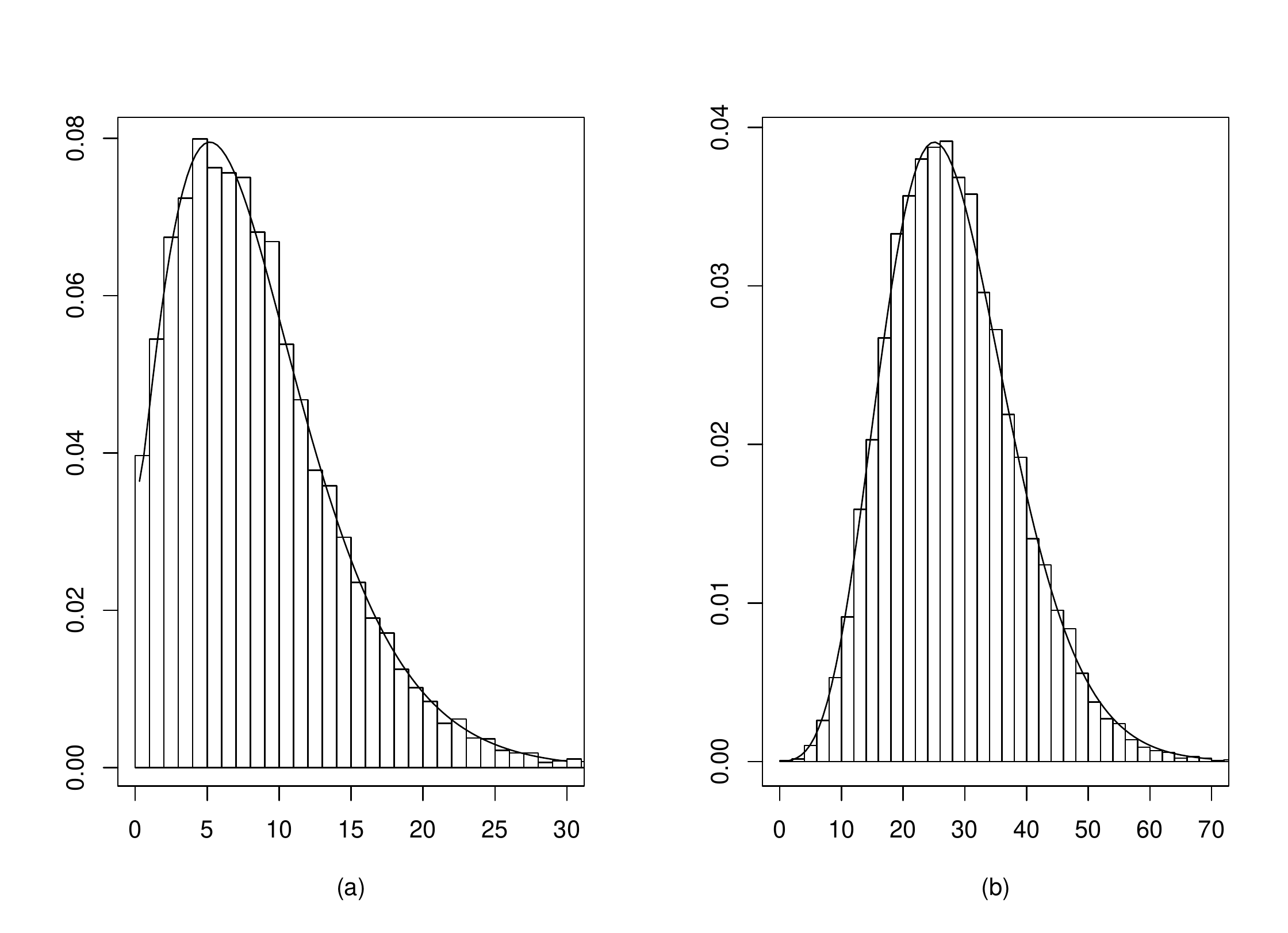}
  \caption{Fit of the distribution of $-2LR$ in the case of nested models, $(g) \subset (h)$ (see section 4.2), by the non-central chi-squared distribution with $q-p$ dof: (a) case of a ``small'' difference of risks (true distribution $f^1$); (b) case of ``moderate'' difference of risks (true distribution $f^2$).}
  \label{Histog}
\end{center}
\end{figure}

\newpage

\begin{figure}[ht]
\begin{center}
\includegraphics[scale=0.8]{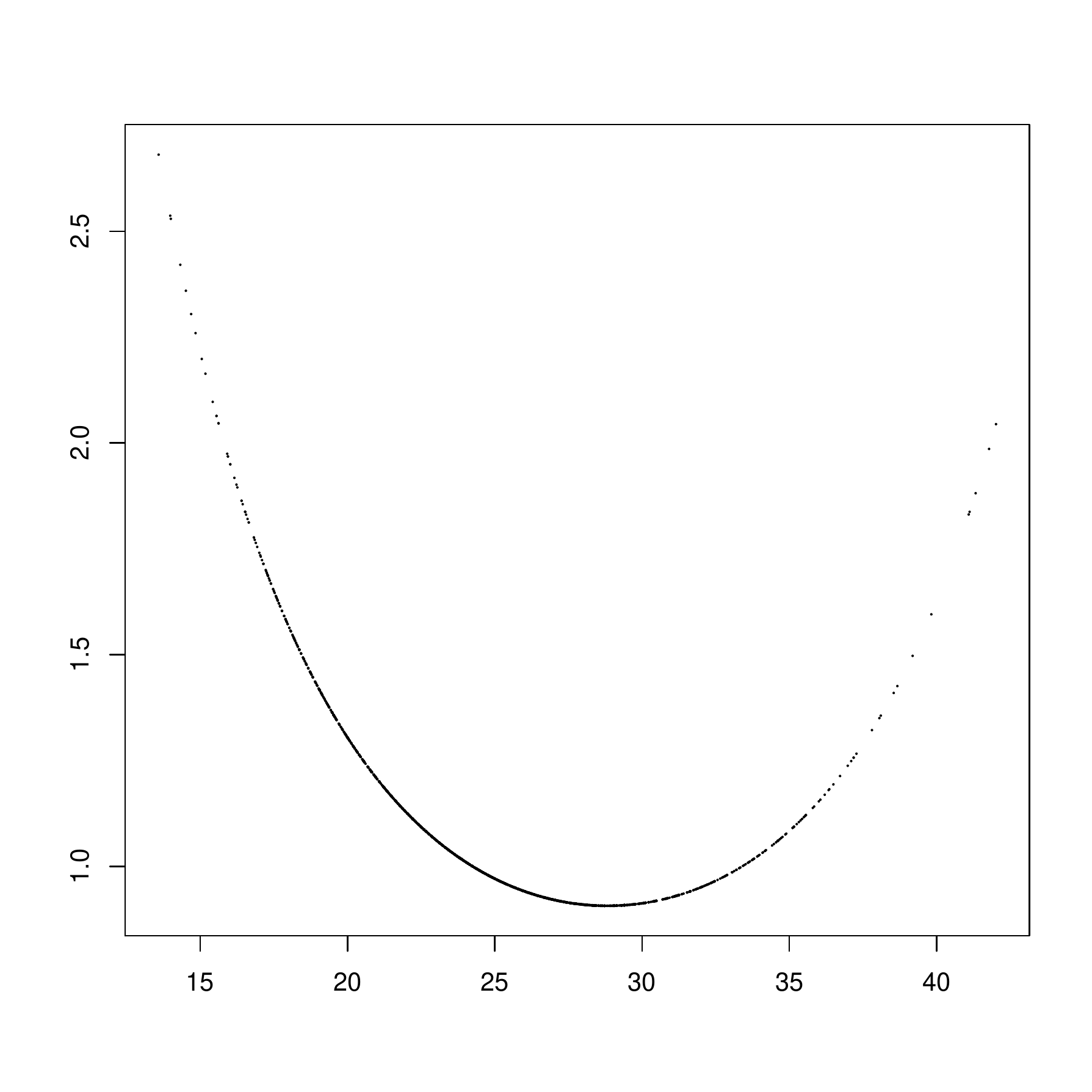}
  \caption{Estimated ``effect'' of the BMI on depression in the quadratic model: odds-ratios with respect to the probability at the median of BMI (24.2); the dots have for abscissas the observed BMI values.}
  \label{Histog}
\end{center}
\end{figure}

\end{document}